\documentclass[%
 reprint,
 amsmath,amssymb,
 aps,prl,noeprint,
floatfix,
]{revtex4-2}

\usepackage{graphicx}
\usepackage{dcolumn}
\usepackage{bm}
\usepackage{braket}
\usepackage{cancel}
\usepackage{hyperref}
\usepackage[compat=1.1.0]{tikz-feynman}
\newcommand{\etal}{{\it et al.}}

\usepackage{hyperref}

\begin{document}

\title{Gallium anomaly and solar neutrinos}
\author{Steven R. Elliott}
\email{Elliotts@lanl.gov}
\affiliation{Physics Division, Los Alamos National Laboratory, Los Alamos NM, USA}

\author{R. G. Hamish Robertson}
\email{rghr@uw.edu}%
\affiliation{Center for Experimental Nuclear Physics and Astrophysics and Department of Physics, University of Washington, Seattle WA, USA}

\date{\today}

\begin{abstract}
The Gallium Anomaly is the disagreement between the measured capture rate of neutrinos by $^{71}$Ga and the rate theoretically expected from the electron capture decay of $^{71}$Ge.  The experimental measurements were performed with intense artificial sources.  In the present work we consider the sun as a neutrino source that can be calibrated via the fundamental, standard-model neutrino-electron elastic scattering reaction in terrestrial detectors. The result is found to be not anomalous, with good agreement between the measured value of $66.1\pm3.1$ SNU from SAGE and Gallex/GNO, and the total measured source strengths of solar sources, $67.5^{+1.4}_{-1.1}$ SNU.
It may be concluded that neither the $^{71}$Ga charged-current cross section nor gallium experiment extraction and counting efficiencies account for the Gallium Anomaly.

\end{abstract}

\maketitle

With the aid of very intense radioactive sources of $^{51}$Cr and $^{37}$Ar, tests have been made of the rate of production of $^{71}$Ge by neutrino interactions on $^{71}$Ga. Initial experiments showed lower rates than expected, with interesting but not conclusive statistical precision.  Recently the BEST experiment \cite{Barinov:2021asz,Barinov:2022wfh} has confirmed this anomaly using a 3.5-MCi $^{51}$Cr source in a specially designed apparatus with two concentric zones of liquid Ga metal.  The apparatus was configured as a short-baseline neutrino oscillation experiment designed to test the possibility that an admixture of sterile neutrinos with electron neutrinos was responsible for the shortfall.  The neutrino capture rates were consistent with no distance dependence.  Together they confirm at more than 4 sigma the low rate indicated by the earlier experiments.

The 5 source experiments and 6 measurements are listed in Table~\ref{tab:gasourceexpts}, as summarized by the BEST collaboration \cite{Barinov:2022wfh}, but with updated cross sections~\cite{Elliott:2023xkb} and ratios.
\begin{table}[htb]
\caption{\label{tab:gasourceexpts}%
Gallium source experiments.
}
\begin{ruledtabular}
\begin{tabular}{lccr}
Experiment & Ratio &  Cross & Ref. \\
& to expected & section & \\
& &  $10^{-45}$ cm$^2$~\cite{Elliott:2023xkb} & \\
\colrule
\vspace{-0.05in}
&&& \\
\vspace{0.05in}
SAGE-Cr & $0.97\pm0.12$ &  $5.69^{+0.28}_{-0.06}$ & \cite{Abdurashitov_PhysRevC.59.2246} \\
\vspace{0.05in}
SAGE-Ar & $0.81\pm0.10$ &  $6.85^{+0.35}_{-0.08}$ & \cite{Abdurashitov:2005tb}\\
\vspace{0.05in}
\sc{Gallex}-Cr1 & $0.97\pm0.11$ &  $5.69^{+0.28}_{-0.06}$ & \cite{Kaether:2010ag}\\
\vspace{0.05in}
\sc{Gallex}-Cr2 & $0.83\pm0.11$ &  $5.69^{+0.28}_{-0.06}$ & \cite{Kaether:2010ag} \\
\vspace{0.05in}
BEST Inner & $0.81\pm0.04$ &  $5.69^{+0.28}_{-0.06}$ & \cite{Barinov:2022wfh}\\
BEST Outer & $0.78\pm0.04$ & $5.69^{+0.28}_{-0.06}$ & \cite{Barinov:2022wfh} \\
\end{tabular}
\end{ruledtabular}
\end{table}
Combining the statistical and experimental systematic uncertainties in quadrature, a weighted average for the 6 measurements can be formed.  When the cross-section uncertainty is then combined in quadrature with that weighted average,
the composite ratio of measurement to expectation is $ 0.817^{+0.028}_{-0.048}$. The uncertainty asymmetry arises from the cross section to the first excited state, which cannot be less than zero.  
This disagreement is termed the `Gallium Anomaly' (GA) and is summarized in~\cite{Elliott:2023xkb,Elliott:2023cvh}.

The basic ground-state to ground-state process is
\begin{eqnarray}
\nu + {^{71}{\rm Ga}}&\rightarrow& {^{71}{\rm Ge}} + e^-.
\end{eqnarray}
The cross section is derived from the electron-capture decay of $^{71}$Ge:
\begin{eqnarray}
e^- + {^{71}{\rm Ge}}&\rightarrow& {^{71}{\rm Ga}} + \nu.
\end{eqnarray}
It is immediately apparent that the two reactions are time-reversed versions of each other.  

The rate for both reactions can be expressed as the product of a matrix element squared and phase-space factors.  Using the same matrix element for both the forward and reverse reactions is an application of `detailed balance' \cite{Perkins}. 

The electron-capture decay of $^{71}$Ge has a precisely determined half-life~\cite{PhysRevC.109.055501,PhysRevC.108.L021602,PhysRevC.31.666} and Q-value~\cite{ALANSSARI20161}, and thus provides the matrix element needed to determine the cross section for the inverse reaction, the capture of a neutrino on $^{71}$Ga to form $^{71}$Ge and an electron in the continuum.  A step in the extraction of the matrix element is factorization of the electron and nuclear wavefunctions.  With factorization the reaction rate becomes proportional simply to the electron density at the nucleus, which can be determined theoretically to percent-level precision by atomic physics calculations.  Recently Cadeddu $\etal$~\cite{Cadeddu2026} have called the factorization assumption into question, suggesting that substantial effects depending on non-separable components of the nuclear and electronic wavefunctions may be responsible for the GA. They propose that this could lead to the cross section for neutrino capture on $^{71}$Ga being 20\% lower than expected from a detailed balance calculation based on factorization. Our long-standing confidence in the use of factorization comes from  experimental measurements of the electron-capture-to-positron ratio in many nuclei.  Generally factorization reproduces these results at the $\sim1$\% level~\cite{Mougeot_2018AppRI.134..225M}, but one exception is known~\cite{Firestone_1978PhRvC..18.2719F} in $^{22}$Na, where the ratio falls below theory by $\sim$10\%. The decay is allowed but hindered by $10^4$, and the deviation is attributed to second-forbidden contributions. Factorization is also used in internal conversion theory, and agreement at the 1\% level is again found~\cite{Hardy:2017exb}. A breakdown of factorization at the 20\% level would thus be very surprising.

The Sun is now an intensively studied and well-quantified source of neutrinos. We here make use of solar neutrinos as a test of the $^{71}$Ga cross section.

There are effectively 5 sources that contribute: pp, pep, $^7$Be, CNO, and $^8$B.  We neglect hep. Only Borexino~\cite{Borexino_PhysRevD.108.102005,Borexino2018,BOREXINO:2020aww} and SNO~\cite{SNO:2011hxd} data are used.  Data from Super-Kamiokande~\cite{Super-Kamiokande:2025hvx,Super-Kamiokande:2023jbt} could also be used, but the spectral analysis is slightly more complicated than for SNO.
Each source has its electron-flavor survival probability $P_{\rm ee}$  determined by oscillations, the solar density profile, and the solar production profile, the latter two of which are solar model dependent.  In our analysis we use the experimental data to avoid model dependence to the extent possible.  The only small model dependence that remains is due to the density dependence of the MSW effect, which requires knowledge of the radial density profile of the production zone for each source.  Terrestrial measurements of the fluxes for $^7$Be, pep, and CNO are obtained from Borexino by elastic scattering (ES).  SNO gives the $^8$B flux via a neutral-current (NC) reaction.  The pp flux contributes strongly and we can determine that flux from the luminosity constraint.

The ES reaction is a fundamental standard-model process~\cite{tHooft:1971ucy}.  The neutrino disintegration of deuterium has been calculated to next-to-next-to-leading order in effective field theory~\cite{Butler:2000zp} and has an estimated uncertainty of 1.3\%~\cite{SNO:2002tuh}.

 The detection rate in the charged-current detectors SAGE~\cite{SAGE:2009eeu}, Gallex/GNO~\cite{Kaether:2010ag},  Cl-Ar~\cite{Cleveland:1994er}, and the SNO charged-current (CC) mode is directly reduced by the electron-neutrino survival probability $P_{ee}$. For the elastic-scattering detectors  the reduction is $a+(1-a)P_{ee}$, where $a$ is an energy-dependent constant $\simeq1/6$ that gives the neutral-current contribution.  For the SNO neutral-current mode there is no reduction.    Rates in the radiochemical detectors and the SNO NC mode are given by an integral over neutrino energies. 

We consider now the fluxes incident on the Ga detectors as measured by  Borexino. The raw rate for the 862-keV neutrino line from $^7$Be is $48.3\pm1.1^{+0.4}_{-0.7}$ per 100 t per day~\cite{Borexino2018}.  The 862-keV line is accompanied by a 384-keV line with a branching ratio of 10.44\%~\cite{Haxton:2025hye}. Converting to the unoscillated flux using $P_{ee} \simeq 0.525$ from the parameters given in~\cite{Esteban:2016qun}, the authors obtain  $4.99\pm 0.11^{+0.06}_{ -0.08} \times 10^9$ cm$^{-2}$ s$^{-1}$ for the total $^7$Be flux.  New results from JUNO~\cite{thejunocollaboration2026measurementneutrinooscillationparameters} favor $P_{ee}=0.555$, for which the flux value becomes $4.79\pm 0.11^{+0.06}_{ -0.08} \times 10^9$ cm$^{-2}$ s$^{-1}$.

The pep line at 1.44 MeV is measured to about  15\% precision by Borexino~\cite{Borexino2018} but there is some dependence on whether one uses the high-Z or low-Z solar model. This dependence is not intrinsic to the pep flux but arises from experimental overlap with the CNO flux in the data.  We therefore fix the pep flux at the calculated fraction $f_{pep}=0.00239(8)$ of the pp flux~\cite{Herrera2023}.  The fraction is very weakly dependent on solar model parameters~\cite{Bahcall:1989ks} and is concordant with  the Borexino results.

Borexino succeeded in detecting the CNO neutrinos in 2020~\cite{BOREXINO:2020aww} and published final results in 2023~\cite{Borexino_PhysRevD.108.102005}. The total  unoscillated CNO flux was found to be $6.7^{+1.2}_{-0.8} \times 10^8$ cm$^{-2}$ s$^{-1}$.  The result was in good agreement with the high-Z model and in more than 2$\sigma$ tension with the low-Z one.

The total unoscillated $^8$B flux is given directly by the SNO neutral-current measurements~\cite{SNO:2011hxd}, and is $5.25\pm 0.16^{+0.11}_{ -0.13} \times 10^6$ cm$^{-2}$ s$^{-1}$.  SNO made use of the $^8$B spectrum obtained by Winter \etal~\cite{Winter:2004kf} but it was recently recommended~\cite{Acharya:2024lke} that the Longfellow \etal~\cite{Longfellow:2023hoj} spectrum be used for future work.  The effect we find is $<0.5$\%, reducing the flux to 
$5.23\pm 0.16^{+0.11}_{ -0.13} \times 10^6$ cm$^{-2}$ s$^{-1}$.

The pp flux has been directly detected in Borexino~\cite{Borexino2018} and found to be $(6.1\pm0.5^{+0.3}_{-0.5})\times10^{10}$ cm$^{-2}$ s$^{-1}$, but a more precise value follows from the luminosity constraint that applies to a star that is in quasi-static equilibrium burning hydrogen to helium. With every helium atom produced, 2 neutrinos and $Q=26.731$ MeV of energy are also produced.  Most of that energy is radiated from the surface of the sun, and about 2\% escapes in the form of neutrinos~\cite{Bahcall:1989ks}.  We use it to evaluate the pp flux given experimental inputs for all the other fluxes.  From the luminosity constraint the pp flux may conveniently be expressed~\cite{Bahcall:1995rs,Heeger:1996se}:
\begin{eqnarray}
    \Phi_{pp}& =& \left[\frac{2I}{Q} - 0.939\Phi_{7}-0.936\Phi_{\rm CNO}  - 0.496 \Phi_8 \right] \nonumber \\
    && \times \left[0.980 (1 - 0.088 f_{pep})\right]^{-1}. 
\end{eqnarray} 
 The irradiance $I = 84.984(280) \times 10^{10}$  MeV cm$^{-2}$ s$^{-1}$ and a table of the coefficients multiplied by $Q/2$ may be found in~\cite{Gonzalez-Garcia:2023kva}.  The coefficient for $\Phi_{\rm CNO}$ is the average of $^{13}$N and $^{15}$O, which are expected to dominate, but all 3 CNO coefficients are very similar, differing at most by 1\% from this.  The pp flux found is $\Phi_{pp}=5.966(27) \times 10^{10}$ cm$^{-2}$ s$^{-1}$.

 The pp spectrum has been calculated by Bahcall~\cite{Bahcall:1997eg}. The Doppler line widths for the $^7$Be and pep sources are negligibly small given the comparatively slow variation of cross sections, but thermal shifts in the line energies are significant for these sources and for the pp neutrinos.  Those shifts are $+3.2$ keV (pp), $+1.29$ keV ($^7$Be 862 keV), $+1.24$ keV ($^7$Be 384 keV), and $+4.5$ keV (pep)~\cite{Haxton:2025hye}. The CNO spectra have the allowed Fermi shape.  

To determine the Ga response to solar neutrinos ($\nu_{\odot}$) we calculate the survival probability of an $\nu_e$ remaining an electron $\nu_e$ ($P_{ee}$)~\cite{BahcallNJP2004}.
\begin{eqnarray}
P_{ee} &=& \left(1 + \frac{[\cos{(2\theta_{12})} - \beta]\cos{(2\theta_{12})}}{\sqrt{[\cos{(2\theta_{12})} - \beta]^2 - \sin^2{(2\theta_{12})}}}\right) \nonumber\\ 
& &\times \frac{\cos^4{(\theta_{13})}}{2}+ \sin^4(\theta_{13}).
\end{eqnarray}
\noindent where $\beta$ is defined by the solar density at the $\nu_e$ creation point, its energy ($E_{\nu}$) and the mass square difference ($\Delta m^2_{12}$),
\begin{eqnarray}
    \beta & = & \frac{2\sqrt{2}G_F \cos^2{(\theta_{13})} n_e E_{\nu}}{\Delta m^2_{12}}   \\
    & = & 0.22 \cos^2{(\theta_{13})}\frac{E_{\nu}}{(1~\mathrm {MeV})}\frac{\mu_e \rho}{(100~\mathrm{g/cm^3})}\frac{7\times10^{-5}}{\Delta m^2_{12}}. \nonumber 
\end{eqnarray}
\noindent where $G_F$ is the Fermi constant, $n_e$ is the electron density, $\mu_{e}$ is the mean electron molecular weight ($\mu_e$ $\approx$ 0.5(1 + X) with $X$ the mass fraction of
hydrogen), $\rho$ is the total density. The values of $n_e$, $\mu_e$, $\rho$ and $X$ are calculated at the $\nu_e$ production point within the Sun.

The neutrino mixing parameter $\sin^2{(2\theta_{13})} = 0.0851\pm0.0024$ for normal ordering is taken from Ref.~\cite{PhysRevLett.130.161802}, and $\Delta m^2_{12} = (7.388\pm0.078)\times10^{-5}$ eV$^2$, and $\sin^2{(\theta_{12})} = 0.3036\pm0.0064$, were taken from Ref.~\cite{thejunocollaboration2026measurementneutrinooscillationparameters}. 

To calculate the survival probability ($P_{ee}$) due to matter oscillations, we used the Herrera-Serenelli MB22p solar model~\cite{Herrera2023}  with photospheric solar abundance data from Ref.~\cite{Magg2022} and fusion data from Ref.~\cite{Acharya:2024lke}. The solar model data included radial solar densities, element fractions and neutrino production distributions. The data were taken from the Zenodo repository~\cite{Herrera2023}. The fluxes from MB22p include those of the CNO branches; $^{13}$N: $3.102\times10^{8}$, $^{15}$O: $2.304\times10^{8}$ and $^{17}$F: $4.699\times10^{6}$ cm$^{-2}$ s$^{-1}$. 

\begin{figure}[htbp]
    \centering
    \includegraphics[width=1.0\linewidth]{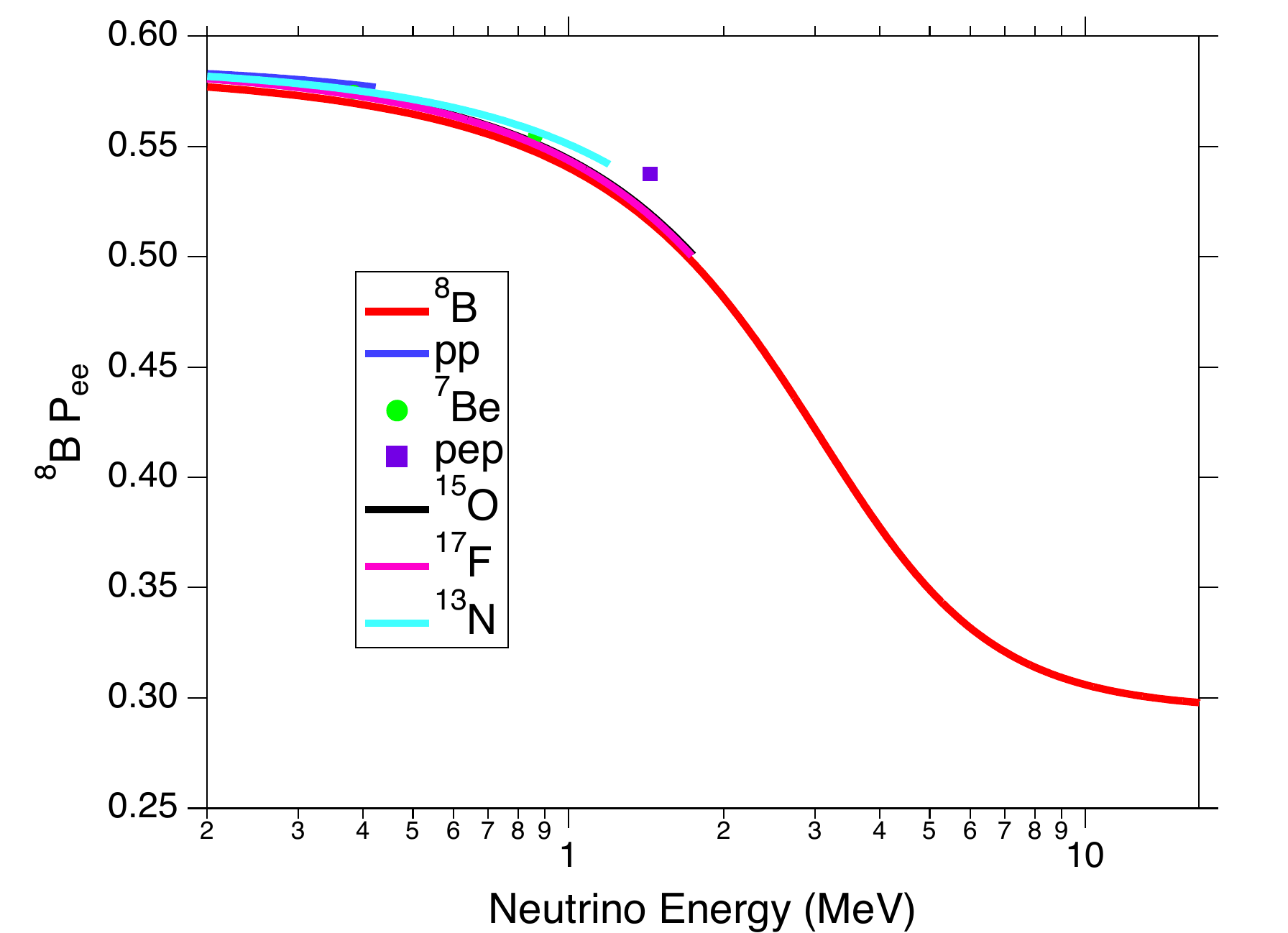}
    \caption{$P_{ee}$ for pp, pep, $^7$Be, CNO, and $^8$B neutrinos using the MB22p solar model and values of the neutrino mixing parameters as described in the text.}
    \label{fig:Pee}
\end{figure}

Figure~\ref{fig:Pee} shows the distributions of the survival probabilities for each of the neutrino species. The pp neutrino energies are all below the critical energy and therefore vacuum oscillations describes them well. $P_{ee}$ for the pp neutrinos is nearly constant between the Ga interaction threshold of 232~keV and the pp spectrum endpoint at 420~keV, varying from 0.588 to 0.577. For the two $^7$Be lines at 0.384 and 0.862~MeV, the $P_{ee}$ values are 0.575 and 0.555 respectively, and the 1.44~MeV pep neutrinos have a $P_{ee}$ of 0.537.

 Detection of the pp, pep, $^7$Be, and CNO fluxes in Ga is dominated by the ground-state transition~\cite{Haxton:2025hye}, the strength of which is precisely determined by detailed balance, although it is also in good agreement with $(p,n)$ reactions.    Only the detection of $^8$B is dominated by transitions to higher excited states, for which the cross sections are determined by $(p,n)$ reactions~\cite{Haxton:2025hye}. 

Bahcall~\cite{Bahcall:1997eg} evaluated the cross sections on $^{71}$Ga for each solar neutrino source.  Haxton and Rule~\cite{Haxton:2025hye} have carried out a new evaluation of these cross sections, which we adopt.  
The cross section for a neutrino of energy $E_\nu$ to produce an electron in the continuum from $^{71}$Ga  is:
\begin{eqnarray}
 \sigma(E_\nu) & =& \frac{G_F^2 \cos^2{\theta_C~ g_A^2}}{\pi}  E_ep_e~{\cal F}(Z_f,E_e) {\mathrm{B}}^{(\nu,e)}_\mathrm{GT} \\
 E_e&=&E_\nu-E_x-Q_{\rm EC}+m_ec^2+\epsilon 
 \end{eqnarray}
 where the electron energy and momentum are $E_e$ and $p_e$, the axial coupling constant $g_A = 1.2764$, the Fermi constant $G_F = 1.1664\times 10^{-5}$ GeV$^{-2}$, the electron mass $m_{e}c^2=511$~keV, and the Cabibbo angle $\cos\theta_C=0.9733$.  The Fermi function ${\cal F}(Z_f,E_e)$, where $Z_f$ is the charge of the daughter nucleus, includes corrections for finite nuclear size and screening. The Gamow-Teller matrix element ${\mathrm{B}}^{(\nu,e)}_\mathrm{GT} $ is tabulated with inclusion of radiative and weak magnetism corrections.  The final state nuclear excitation energy is $E_x$ and $\epsilon$ is a small correction for atomic rearrangement.  In $^{71}$Ge, Bahcall~\cite{Bahcall:1997eg} finds $\epsilon=-90$ eV.  The electron capture Q-value is $Q_{\rm EC}=232.443(93)$ keV~\cite{ALANSSARI20161}.  To determine the ground-state ${\mathrm{B}}^{(\nu,e)}_\mathrm{GT}$, Haxton and Rule use a weighted average of 3 measurements~\cite{PhysRevC.109.055501,PhysRevC.108.L021602,PhysRevC.31.666} of the $^{71}$Ge half-life, 11.465(8) d. 
 
\begin{figure}[htbp]
    \centering
    \includegraphics[width=1.0\linewidth]{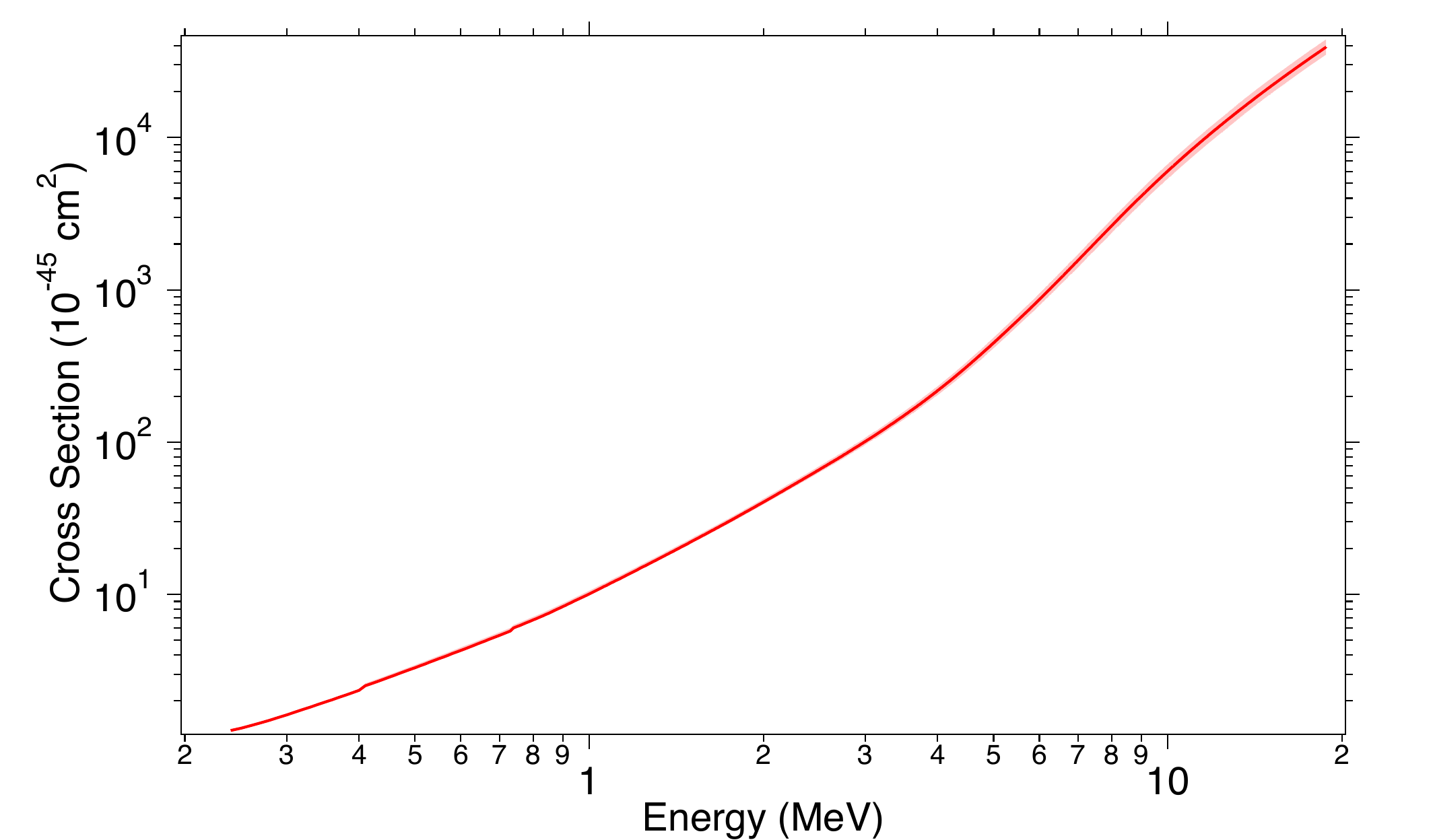}
    \caption{The $^{71}$Ga($\nu_e$,e$^-$)$^{71}$Ge cross section as calculated by Haxton and Rule~\cite{Haxton:2025hye}. The line is the value and the shaded region is their estimate of the uncertainty.}
    \label{fig:CrossSection}
\end{figure}

 Figure~\ref{fig:CrossSection} shows the results.
The uncertainties in Fig.~\ref{fig:CrossSection} are not symmetric, especially at  energies where the cross section is sensitive to transitions to the first excited state in $^{71}$Ge, for which only an upper bound is known from experiment. At those energies, the upper uncertainty is about 5\% and a bit less than 1\% for pp neutrinos. At the highest energies, the uncertainty is about 10\%.

For each solar electron neutrino source $i$, we calculate the detection rate for neutrinos with an oscillated spectrum ($R_{osc}^i$)
\begin{eqnarray}
    R_{osc}^i &=& \int_{Q_{EC}}^{E^i_0} \sigma(E_{\nu}) \frac{d\Phi_i}{dE_{\nu}} P_{ee}(E_{\nu}) dE_{\nu}  \\ 
    \left<\sigma_{e,i}\right>&=& \frac{R_{osc}^i}{\Phi_i},
\end{eqnarray}
where $\frac{d\Phi_i}{dE_{\nu}}$ is the neutrino energy spectrum with endpoint energy ($E^i_0$). Borexino and SNO report their results in terms of the flux of neutrinos $\Phi_i$ emitted by a specific solar source.  Those determinations may then be applied to the calculation of the detection rate in a gallium detector by means of the effective cross section $\left<\sigma_{e,i}\right>$.

\begin{table}[!hbt]
    \centering
    \caption{Gallium detection rates deduced from the solar neutrino sources, their strengths being experimentally determined from $(\nu,e)$ elastic scattering and d$(\nu,\nu^\prime)$pn neutral-current reactions.  Electron-flavor survival has been determined from Daya Bay~\cite{PhysRevLett.130.161802} and JUNO~\cite{thejunocollaboration2026measurementneutrinooscillationparameters} measurements. MSW conversion depends on the solar density profile; solar models are not otherwise used.  Cross sections are from Haxton and Rule~\cite{Haxton:2025hye}. The last column gives the prediction of the MB22p standard solar model~\cite{Herrera2023}.}
    \begin{tabular}{lrcrr}
    \hline\hline
    Reaction & $\left<\sigma_{e,i}\right>$   &  Measured flux                       &  Deduced Ga  & MB22p \\
            & ($10^{-46}$cm$^2$)            & (cm$^{-2}$ s$^{-1}$)                  & Rate (SNU)     & (SNU)  \\
    \hline
    \rule{0pt}{2.5ex}pp  & 6.70   & $5.966\times10^{10}$                & 40.00         & 39.89 \\
    pep		& 109.00                       & $1.43 \times10^{8}$                       & 1.55       & 1.55 \\ 
    $^7$Be  & 39.14                         & ${4.79\times10^9}$   & 18.75          & 19.08 \\ 
    $^8$B	& 7981.87                         & $5.23\times10^6$      & 4.17          & 4.05 \\ 
    CNO		&  44.77                           & $6.7 \times10^{8}$    & 3.00          & 2.44  \\
\hline
    \rule{0pt}{2.5ex}Total   &  &   & $67.47^{+1.42}_{-1.11}$  & 67.01 \\
    \hline
    \multicolumn{3}{l}{\rule{0pt}{2.5ex}SAGE~\cite{SAGE:2009eeu}} & $65.4^{+3.1\ +2.6}_{-3.4\ -2.8}$ &  \\
    \multicolumn{3}{l}{\rule{0pt}{2.5ex}Gallex/GNO~\cite{Kaether:2010ag}} & $73.4^{+7.1}_{-7.3}$ &  \\
    \multicolumn{3}{l}{\rule{0pt}{2.5ex}Weighted Average~\cite{SAGE:2009eeu}} & $66.1\pm 3.1$ &  \\
    \hline\hline
    \end{tabular}
    \label{tab:DetectionRates}
\end{table}
The resulting cross sections, experimental inputs, the corresponding SNU rates in Ga (captures per $10^{36}$ target atoms per second), and (for comparison) the predicted SNU rates from the MB22p standard solar model of Herrera and Serenelli~\cite{Herrera2023} are given in Table~\ref{tab:DetectionRates}.  The uncertainty contribution of each input to the calculated rate of $^{71}$Ga neutrino interactions is given in Table~\ref{tab:uncert}.
\begin{table}[hbt]
    \centering
    \caption{Standard-deviation uncertainties in the $^{71}$Ga detection rate.  }
    \begin{tabular}{lr}
    \hline\hline
    Input parameter & Contribution (SNU) \\
    \hline
    $\sin{\theta_{12}}$ & $\pm0.64$ \\ 
    $\Delta m_{12}^2$ & $\pm0.03$ \\
    $\sin{\theta_{13}}$ & $\pm0.08$ \\ 
    Irradiance   &  $\pm0.15$ \\
    $f_{pep}$   &   $\pm0.05$   \\
    $^7$Be flux     & $\pm0.43$ \\
    $^8$B flux     & $\pm0.16$ \\
    CNO flux     & $\pm0.43$ \\
    pp cross section & +0.36, -0.32 \\
    pep+$^7$Be+CNO cross section    & +0.90, -0.28  \\
    $^8$B cross section & +0.44, -0.44  \\
    Solar models  & $\pm0.19$ \\
    \hline
    Total uncertainty & +1.42, -1.11  \\
    \hline\hline
    \end{tabular}
    \label{tab:uncert}
\end{table}
\begin{figure}[!hbt]
    \centering
    \includegraphics[width=3.4in]{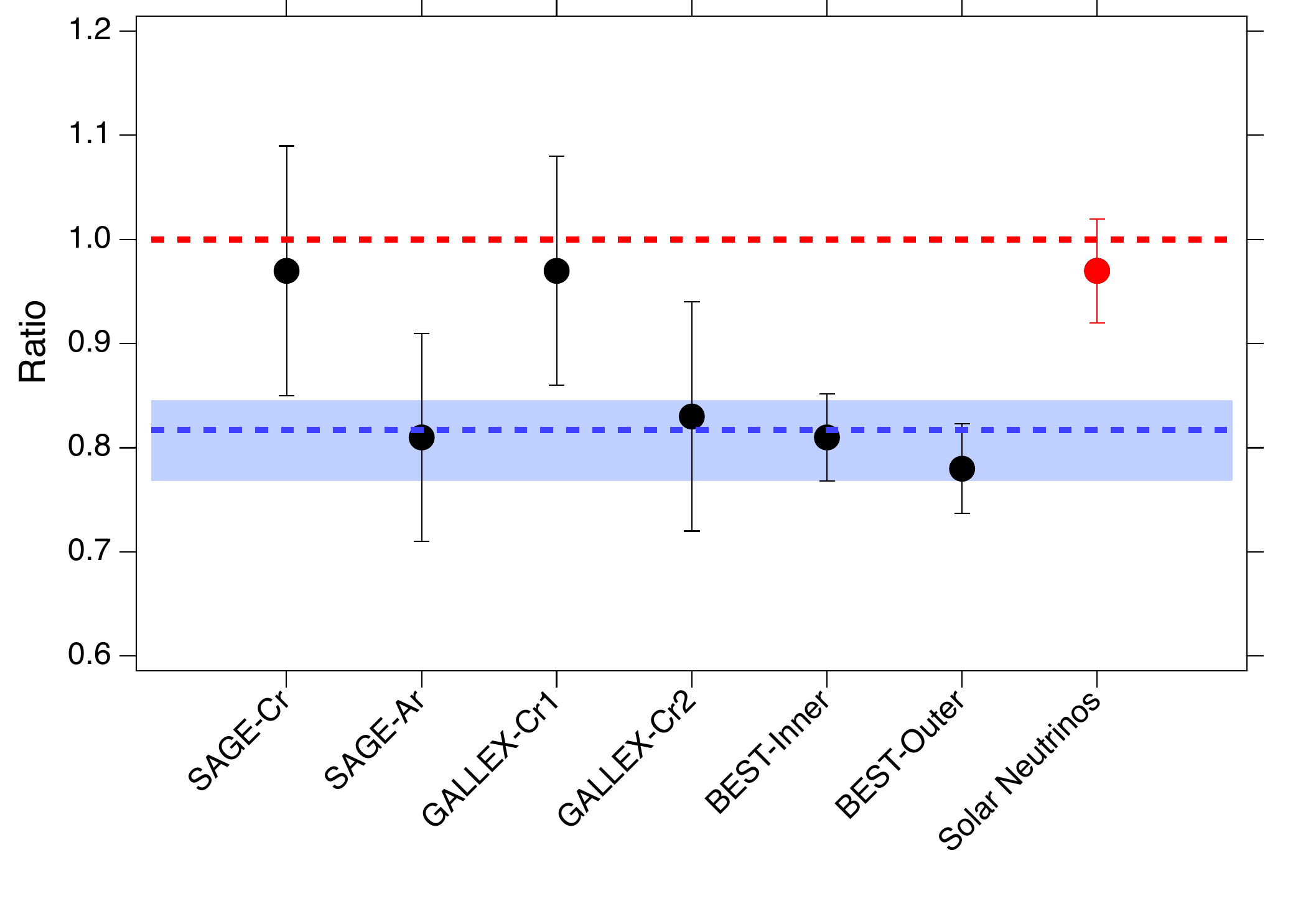}
    \caption{Ratio of detected rates to expected rates for neutrino capture on $^{71}$Ga. The black points are from Ref.~\cite{Barinov:2022wfh} and the red point is from this work. The blue band is the assigned uncertainty range for the average of the black points~\cite{Elliott:2023xkb}.}
    \label{fig:table&plot}
\end{figure}
In summary, using the neutrino fluxes measured by neutrino-electron elastic scattering and by neutral-current disintegration of deuterium, we estimated the predicted measurement for the Ga experiments. The result $67.5^{+1.4}_{-1.1}$ SNU agrees very well with the combined Gallex/GNO - SAGE result of $66.1\pm3.1$ SNU~\cite{SAGE:2009eeu}. The ratio of these, plotted in Fig.~\ref{fig:table&plot}, is $0.980^{+0.050}_{-0.049}$. 
More than 90\% of the rate calculated is from cross sections derived from electron capture and detailed balance, the remainder coming from $(p,n)$ reaction data. The gallium anomaly then apparently does not arise from an error such as the breakdown of factorization suggested by Cadeddu \etal~\cite{Cadeddu2026}, nor indeed is there indication of any error in the $^{71}$Ga cross section.

The result also indicates that the stated experimental extraction and counting efficiencies for the solar neutrino runs by SAGE, Gallex and BEST are accurate. Since the procedures are similar, it is likely that the source experiment efficiencies are accurate as well. 

The source of the gallium anomaly evidently lies elsewhere.

This material is based upon work supported by the U.S.~Department of Energy, Office of Science, Office of Nuclear Physics under Federal Prime Agreement LANLEM78 (SRE) and  Award Number DE-FG02-97ER41020 (RGHR). We thank Evan Rule for providing theoretical cross section data for Ga.

\bibliography{gallium}

\begin{thebibliography}{42}%
\makeatletter
\providecommand \@ifxundefined [1]{%
 \@ifx{#1\undefined}
}%
\providecommand \@ifnum [1]{%
 \ifnum #1\expandafter \@firstoftwo
 \else \expandafter \@secondoftwo
 \fi
}%
\providecommand \@ifx [1]{%
 \ifx #1\expandafter \@firstoftwo
 \else \expandafter \@secondoftwo
 \fi
}%
\providecommand \natexlab [1]{#1}%
\providecommand \enquote  [1]{``#1''}%
\providecommand \bibnamefont  [1]{#1}%
\providecommand \bibfnamefont [1]{#1}%
\providecommand \citenamefont [1]{#1}%
\providecommand \href@noop [0]{\@secondoftwo}%
\providecommand \href [0]{\begingroup \@sanitize@url \@href}%
\providecommand \@href[1]{\@@startlink{#1}\@@href}%
\providecommand \@@href[1]{\endgroup#1\@@endlink}%
\providecommand \@sanitize@url [0]{\catcode `\\12\catcode `\$12\catcode `\&12\catcode `\#12\catcode `\^12\catcode `\_12\catcode `\%12\relax}%
\providecommand \@@startlink[1]{}%
\providecommand \@@endlink[0]{}%
\providecommand \url  [0]{\begingroup\@sanitize@url \@url }%
\providecommand \@url [1]{\endgroup\@href {#1}{\urlprefix }}%
\providecommand \urlprefix  [0]{URL }%
\providecommand \Eprint [0]{\href }%
\providecommand \doibase [0]{http://dx.doi.org/}%
\providecommand \selectlanguage [0]{\@gobble}%
\providecommand \bibinfo  [0]{\@secondoftwo}%
\providecommand \bibfield  [0]{\@secondoftwo}%
\providecommand \translation [1]{[#1]}%
\providecommand \BibitemOpen [0]{}%
\providecommand \bibitemStop [0]{}%
\providecommand \bibitemNoStop [0]{.\EOS\space}%
\providecommand \EOS [0]{\spacefactor3000\relax}%
\providecommand \BibitemShut  [1]{\csname bibitem#1\endcsname}%
\let\auto@bib@innerbib\@empty
\bibitem [{\citenamefont {Barinov}\ \emph {et~al.}(2022{\natexlab{a}})\citenamefont {Barinov}, \citenamefont {Cleveland}, \citenamefont {Danshin}, \citenamefont {Ejiri}, \citenamefont {Elliott}, \citenamefont {Frekers}, \citenamefont {Gavrin}, \citenamefont {Gorbachev}, \citenamefont {Gorbunov}, \citenamefont {Haxton}, \citenamefont {Ibragimova}, \citenamefont {Kim}, \citenamefont {Kozlova}, \citenamefont {Kravchuk}, \citenamefont {Kuzminov}, \citenamefont {Lubsandorzhiev}, \citenamefont {Malyshkin}, \citenamefont {Massarczyk}, \citenamefont {Matveev}, \citenamefont {Mirmov}, \citenamefont {Nico}, \citenamefont {Petelin}, \citenamefont {Robertson}, \citenamefont {Sinclair}, \citenamefont {Shikhin}, \citenamefont {Tarasov}, \citenamefont {Trubnikov}, \citenamefont {Veretenkin}, \citenamefont {Wilkerson},\ and\ \citenamefont {Zvir}}]{Barinov:2021asz}%
  \BibitemOpen
  \bibfield  {author} {\bibinfo {author} {\bibfnamefont {V.~V.}\ \bibnamefont {Barinov}}, \bibinfo {author} {\bibfnamefont {B.~T.}\ \bibnamefont {Cleveland}}, \bibinfo {author} {\bibfnamefont {S.~N.}\ \bibnamefont {Danshin}}, \bibinfo {author} {\bibfnamefont {H.}~\bibnamefont {Ejiri}}, \bibinfo {author} {\bibfnamefont {S.~R.}\ \bibnamefont {Elliott}}, \bibinfo {author} {\bibfnamefont {D.}~\bibnamefont {Frekers}}, \bibinfo {author} {\bibfnamefont {V.~N.}\ \bibnamefont {Gavrin}}, \bibinfo {author} {\bibfnamefont {V.~V.}\ \bibnamefont {Gorbachev}}, \bibinfo {author} {\bibfnamefont {D.~S.}\ \bibnamefont {Gorbunov}}, \bibinfo {author} {\bibfnamefont {W.~C.}\ \bibnamefont {Haxton}}, \bibinfo {author} {\bibfnamefont {T.~V.}\ \bibnamefont {Ibragimova}}, \bibinfo {author} {\bibfnamefont {I.}~\bibnamefont {Kim}}, \bibinfo {author} {\bibfnamefont {Y.~P.}\ \bibnamefont {Kozlova}}, \bibinfo {author} {\bibfnamefont {L.~V.}\ \bibnamefont {Kravchuk}}, \bibinfo {author} {\bibfnamefont {V.~V.}\ \bibnamefont {Kuzminov}},
  \bibinfo {author} {\bibfnamefont {B.~K.}\ \bibnamefont {Lubsandorzhiev}}, \bibinfo {author} {\bibfnamefont {Y.~M.}\ \bibnamefont {Malyshkin}}, \bibinfo {author} {\bibfnamefont {R.}~\bibnamefont {Massarczyk}}, \bibinfo {author} {\bibfnamefont {V.~A.}\ \bibnamefont {Matveev}}, \bibinfo {author} {\bibfnamefont {I.~N.}\ \bibnamefont {Mirmov}}, \bibinfo {author} {\bibfnamefont {J.~S.}\ \bibnamefont {Nico}}, \bibinfo {author} {\bibfnamefont {A.~L.}\ \bibnamefont {Petelin}}, \bibinfo {author} {\bibfnamefont {R.~G.~H.}\ \bibnamefont {Robertson}}, \bibinfo {author} {\bibfnamefont {D.}~\bibnamefont {Sinclair}}, \bibinfo {author} {\bibfnamefont {A.~A.}\ \bibnamefont {Shikhin}}, \bibinfo {author} {\bibfnamefont {V.~A.}\ \bibnamefont {Tarasov}}, \bibinfo {author} {\bibfnamefont {G.~V.}\ \bibnamefont {Trubnikov}}, \bibinfo {author} {\bibfnamefont {E.~P.}\ \bibnamefont {Veretenkin}}, \bibinfo {author} {\bibfnamefont {J.~F.}\ \bibnamefont {Wilkerson}}, \ and\ \bibinfo {author} {\bibfnamefont {A.~I.}\ \bibnamefont {Zvir}},\
  }\href {\doibase 10.1103/PhysRevLett.128.232501} {\bibfield  {journal} {\bibinfo  {journal} {Phys. Rev. Lett.}\ }\textbf {\bibinfo {volume} {128}},\ \bibinfo {pages} {232501} (\bibinfo {year} {2022}{\natexlab{a}})},\ \Eprint {http://arxiv.org/abs/2109.11482} {arXiv:2109.11482} \BibitemShut {NoStop}%
\bibitem [{\citenamefont {Barinov}\ \emph {et~al.}(2022{\natexlab{b}})\citenamefont {Barinov}, \citenamefont {Danshin}, \citenamefont {Gavrin}, \citenamefont {Gorbachev}, \citenamefont {Gorbunov}, \citenamefont {Ibragimova}, \citenamefont {Kozlova}, \citenamefont {Kravchuk}, \citenamefont {Kuzminov}, \citenamefont {Lubsandorzhiev}, \citenamefont {Malyshkin}, \citenamefont {Mirmov}, \citenamefont {Shikhin}, \citenamefont {Veretenkin}, \citenamefont {Cleveland}, \citenamefont {Ejiri}, \citenamefont {Elliott}, \citenamefont {Kim}, \citenamefont {Massarczyk}, \citenamefont {Frekers}, \citenamefont {Haxton}, \citenamefont {Matveev}, \citenamefont {Trubnikov}, \citenamefont {Nico}, \citenamefont {Petelin}, \citenamefont {Tarasov}, \citenamefont {Zvir}, \citenamefont {Robertson}, \citenamefont {Sinclair},\ and\ \citenamefont {Wilkerson}}]{Barinov:2022wfh}%
  \BibitemOpen
  \bibfield  {author} {\bibinfo {author} {\bibfnamefont {V.~V.}\ \bibnamefont {Barinov}}, \bibinfo {author} {\bibfnamefont {S.~N.}\ \bibnamefont {Danshin}}, \bibinfo {author} {\bibfnamefont {V.~N.}\ \bibnamefont {Gavrin}}, \bibinfo {author} {\bibfnamefont {V.~V.}\ \bibnamefont {Gorbachev}}, \bibinfo {author} {\bibfnamefont {D.~S.}\ \bibnamefont {Gorbunov}}, \bibinfo {author} {\bibfnamefont {T.~V.}\ \bibnamefont {Ibragimova}}, \bibinfo {author} {\bibfnamefont {Y.~P.}\ \bibnamefont {Kozlova}}, \bibinfo {author} {\bibfnamefont {L.~V.}\ \bibnamefont {Kravchuk}}, \bibinfo {author} {\bibfnamefont {V.~V.}\ \bibnamefont {Kuzminov}}, \bibinfo {author} {\bibfnamefont {B.~K.}\ \bibnamefont {Lubsandorzhiev}}, \bibinfo {author} {\bibfnamefont {Y.~M.}\ \bibnamefont {Malyshkin}}, \bibinfo {author} {\bibfnamefont {I.~N.}\ \bibnamefont {Mirmov}}, \bibinfo {author} {\bibfnamefont {A.~A.}\ \bibnamefont {Shikhin}}, \bibinfo {author} {\bibfnamefont {E.~P.}\ \bibnamefont {Veretenkin}}, \bibinfo {author} {\bibfnamefont {B.~T.}\
  \bibnamefont {Cleveland}}, \bibinfo {author} {\bibfnamefont {H.}~\bibnamefont {Ejiri}}, \bibinfo {author} {\bibfnamefont {S.~R.}\ \bibnamefont {Elliott}}, \bibinfo {author} {\bibfnamefont {I.}~\bibnamefont {Kim}}, \bibinfo {author} {\bibfnamefont {R.}~\bibnamefont {Massarczyk}}, \bibinfo {author} {\bibfnamefont {D.}~\bibnamefont {Frekers}}, \bibinfo {author} {\bibfnamefont {W.~C.}\ \bibnamefont {Haxton}}, \bibinfo {author} {\bibfnamefont {V.~A.}\ \bibnamefont {Matveev}}, \bibinfo {author} {\bibfnamefont {G.~V.}\ \bibnamefont {Trubnikov}}, \bibinfo {author} {\bibfnamefont {J.~S.}\ \bibnamefont {Nico}}, \bibinfo {author} {\bibfnamefont {A.~L.}\ \bibnamefont {Petelin}}, \bibinfo {author} {\bibfnamefont {V.~A.}\ \bibnamefont {Tarasov}}, \bibinfo {author} {\bibfnamefont {A.~I.}\ \bibnamefont {Zvir}}, \bibinfo {author} {\bibfnamefont {R.~G.~H.}\ \bibnamefont {Robertson}}, \bibinfo {author} {\bibfnamefont {D.}~\bibnamefont {Sinclair}}, \ and\ \bibinfo {author} {\bibfnamefont {J.~F.}\ \bibnamefont {Wilkerson}},\
  }\href {\doibase 10.1103/PhysRevC.105.065502} {\bibfield  {journal} {\bibinfo  {journal} {Phys. Rev. C}\ }\textbf {\bibinfo {volume} {105}},\ \bibinfo {pages} {065502} (\bibinfo {year} {2022}{\natexlab{b}})},\ \Eprint {http://arxiv.org/abs/2201.07364} {arXiv:2201.07364} \BibitemShut {NoStop}%
\bibitem [{\citenamefont {Elliott}\ \emph {et~al.}(2023)\citenamefont {Elliott}, \citenamefont {Gavrin}, \citenamefont {Haxton}, \citenamefont {Ibragimova},\ and\ \citenamefont {Rule}}]{Elliott:2023xkb}%
  \BibitemOpen
  \bibfield  {author} {\bibinfo {author} {\bibfnamefont {S.~R.}\ \bibnamefont {Elliott}}, \bibinfo {author} {\bibfnamefont {V.~N.}\ \bibnamefont {Gavrin}}, \bibinfo {author} {\bibfnamefont {W.~C.}\ \bibnamefont {Haxton}}, \bibinfo {author} {\bibfnamefont {T.~V.}\ \bibnamefont {Ibragimova}}, \ and\ \bibinfo {author} {\bibfnamefont {E.~J.}\ \bibnamefont {Rule}},\ }\href {\doibase 10.1103/PhysRevC.108.035502} {\bibfield  {journal} {\bibinfo  {journal} {Phys. Rev. C}\ }\textbf {\bibinfo {volume} {108}},\ \bibinfo {pages} {035502} (\bibinfo {year} {2023})},\ \Eprint {http://arxiv.org/abs/2303.13623v3} {arXiv:2303.13623v3} \BibitemShut {NoStop}%
\bibitem [{\citenamefont {Abdurashitov}\ \emph {et~al.}(1999)\citenamefont {Abdurashitov}, \citenamefont {Gavrin}, \citenamefont {Girin}, \citenamefont {Gorbachev}, \citenamefont {Ibragimova}, \citenamefont {Kalikhov}, \citenamefont {Khairnasov}, \citenamefont {Knodel}, \citenamefont {Kornoukhov}, \citenamefont {Mirmov}, \citenamefont {Shikhin}, \citenamefont {Veretenkin}, \citenamefont {Vermul}, \citenamefont {Yants}, \citenamefont {Zatsepin}, \citenamefont {Khomyakov}, \citenamefont {Zvonarev}, \citenamefont {Bowles}, \citenamefont {Nico}, \citenamefont {Teasdale}, \citenamefont {Wark}, \citenamefont {Cherry}, \citenamefont {Karaulov}, \citenamefont {Levitin}, \citenamefont {Maev}, \citenamefont {Nazarenko}, \citenamefont {Shkol'nik}, \citenamefont {Skorikov}, \citenamefont {Cleveland}, \citenamefont {Daily}, \citenamefont {Davis}, \citenamefont {Lande}, \citenamefont {Lee}, \citenamefont {Wildenhain}, \citenamefont {Elliott},\ and\ \citenamefont {Wilkerson}}]{Abdurashitov_PhysRevC.59.2246}%
  \BibitemOpen
  \bibfield  {author} {\bibinfo {author} {\bibfnamefont {J.~N.}\ \bibnamefont {Abdurashitov}}, \bibinfo {author} {\bibfnamefont {V.~N.}\ \bibnamefont {Gavrin}}, \bibinfo {author} {\bibfnamefont {S.~V.}\ \bibnamefont {Girin}}, \bibinfo {author} {\bibfnamefont {V.~V.}\ \bibnamefont {Gorbachev}}, \bibinfo {author} {\bibfnamefont {T.~V.}\ \bibnamefont {Ibragimova}}, \bibinfo {author} {\bibfnamefont {A.~V.}\ \bibnamefont {Kalikhov}}, \bibinfo {author} {\bibfnamefont {N.~G.}\ \bibnamefont {Khairnasov}}, \bibinfo {author} {\bibfnamefont {T.~V.}\ \bibnamefont {Knodel}}, \bibinfo {author} {\bibfnamefont {V.~N.}\ \bibnamefont {Kornoukhov}}, \bibinfo {author} {\bibfnamefont {I.~N.}\ \bibnamefont {Mirmov}}, \bibinfo {author} {\bibfnamefont {A.~A.}\ \bibnamefont {Shikhin}}, \bibinfo {author} {\bibfnamefont {E.~P.}\ \bibnamefont {Veretenkin}}, \bibinfo {author} {\bibfnamefont {V.~M.}\ \bibnamefont {Vermul}}, \bibinfo {author} {\bibfnamefont {V.~E.}\ \bibnamefont {Yants}}, \bibinfo {author} {\bibfnamefont {G.~T.}\
  \bibnamefont {Zatsepin}}, \bibinfo {author} {\bibfnamefont {Y.~S.}\ \bibnamefont {Khomyakov}}, \bibinfo {author} {\bibfnamefont {A.~V.}\ \bibnamefont {Zvonarev}}, \bibinfo {author} {\bibfnamefont {T.~J.}\ \bibnamefont {Bowles}}, \bibinfo {author} {\bibfnamefont {J.~S.}\ \bibnamefont {Nico}}, \bibinfo {author} {\bibfnamefont {W.~A.}\ \bibnamefont {Teasdale}}, \bibinfo {author} {\bibfnamefont {D.~L.}\ \bibnamefont {Wark}}, \bibinfo {author} {\bibfnamefont {M.~L.}\ \bibnamefont {Cherry}}, \bibinfo {author} {\bibfnamefont {V.~N.}\ \bibnamefont {Karaulov}}, \bibinfo {author} {\bibfnamefont {V.~L.}\ \bibnamefont {Levitin}}, \bibinfo {author} {\bibfnamefont {V.~I.}\ \bibnamefont {Maev}}, \bibinfo {author} {\bibfnamefont {P.~I.}\ \bibnamefont {Nazarenko}}, \bibinfo {author} {\bibfnamefont {V.~S.}\ \bibnamefont {Shkol'nik}}, \bibinfo {author} {\bibfnamefont {N.~V.}\ \bibnamefont {Skorikov}}, \bibinfo {author} {\bibfnamefont {B.~T.}\ \bibnamefont {Cleveland}}, \bibinfo {author} {\bibfnamefont {T.}~\bibnamefont
  {Daily}}, \bibinfo {author} {\bibfnamefont {R.}~\bibnamefont {Davis}}, \bibinfo {author} {\bibfnamefont {K.}~\bibnamefont {Lande}}, \bibinfo {author} {\bibfnamefont {C.~K.}\ \bibnamefont {Lee}}, \bibinfo {author} {\bibfnamefont {P.~S.}\ \bibnamefont {Wildenhain}}, \bibinfo {author} {\bibfnamefont {S.~R.}\ \bibnamefont {Elliott}}, \ and\ \bibinfo {author} {\bibfnamefont {J.~F.}\ \bibnamefont {Wilkerson}} (\bibinfo {collaboration} {The SAGE Collaboration}),\ }\href {\doibase 10.1103/PhysRevC.59.2246} {\bibfield  {journal} {\bibinfo  {journal} {Phys. Rev. C}\ }\textbf {\bibinfo {volume} {59}},\ \bibinfo {pages} {2246} (\bibinfo {year} {1999})}\BibitemShut {NoStop}%
\bibitem [{\citenamefont {Abdurashitov}\ \emph {et~al.}(2006)\citenamefont {Abdurashitov} \emph {et~al.}}]{Abdurashitov:2005tb}%
  \BibitemOpen
  \bibfield  {author} {\bibinfo {author} {\bibfnamefont {J.~N.}\ \bibnamefont {Abdurashitov}} \emph {et~al.},\ }\href {\doibase 10.1103/PhysRevC.73.045805} {\bibfield  {journal} {\bibinfo  {journal} {Phys. Rev. C}\ }\textbf {\bibinfo {volume} {73}},\ \bibinfo {pages} {045805} (\bibinfo {year} {2006})},\ \Eprint {http://arxiv.org/abs/nucl-ex/0512041} {arXiv:nucl-ex/0512041} \BibitemShut {NoStop}%
\bibitem [{\citenamefont {Kaether}\ \emph {et~al.}(2010)\citenamefont {Kaether}, \citenamefont {Hampel}, \citenamefont {Heusser}, \citenamefont {Kiko},\ and\ \citenamefont {Kirsten}}]{Kaether:2010ag}%
  \BibitemOpen
  \bibfield  {author} {\bibinfo {author} {\bibfnamefont {F.}~\bibnamefont {Kaether}}, \bibinfo {author} {\bibfnamefont {W.}~\bibnamefont {Hampel}}, \bibinfo {author} {\bibfnamefont {G.}~\bibnamefont {Heusser}}, \bibinfo {author} {\bibfnamefont {J.}~\bibnamefont {Kiko}}, \ and\ \bibinfo {author} {\bibfnamefont {T.}~\bibnamefont {Kirsten}},\ }\href {\doibase 10.1016/j.physletb.2010.01.030} {\bibfield  {journal} {\bibinfo  {journal} {Phys. Lett. B}\ }\textbf {\bibinfo {volume} {685}},\ \bibinfo {pages} {47} (\bibinfo {year} {2010})},\ \Eprint {http://arxiv.org/abs/1001.2731} {arXiv:1001.2731 [hep-ex]} \BibitemShut {NoStop}%
\bibitem [{\citenamefont {Elliott}\ \emph {et~al.}(2024)\citenamefont {Elliott}, \citenamefont {Gavrin},\ and\ \citenamefont {Haxton}}]{Elliott:2023cvh}%
  \BibitemOpen
  \bibfield  {author} {\bibinfo {author} {\bibfnamefont {S.}~\bibnamefont {Elliott}}, \bibinfo {author} {\bibfnamefont {V.}~\bibnamefont {Gavrin}}, \ and\ \bibinfo {author} {\bibfnamefont {W.}~\bibnamefont {Haxton}},\ }\href {\doibase https://doi.org/10.1016/j.ppnp.2023.104082} {\bibfield  {journal} {\bibinfo  {journal} {Progress in Particle and Nuclear Physics}\ }\textbf {\bibinfo {volume} {134}},\ \bibinfo {pages} {104082} (\bibinfo {year} {2024})},\ \Eprint {http://arxiv.org/abs/2306.03299} {arXiv:2306.03299} \BibitemShut {NoStop}%
\bibitem [{\citenamefont {Perkins}(1987)}]{Perkins}%
  \BibitemOpen
  \bibfield  {author} {\bibinfo {author} {\bibfnamefont {D.~H.}\ \bibnamefont {Perkins}},\ }\href@noop {} {\emph {\bibinfo {title} {{Introduction to High Energy Physics}}}},\ \bibinfo {edition} {3rd}\ ed.\ (\bibinfo  {publisher} {Addison Wesley},\ \bibinfo {year} {1987})\BibitemShut {NoStop}%
\bibitem [{\citenamefont {Norman}\ \emph {et~al.}(2024)\citenamefont {Norman}, \citenamefont {Drobizhev}, \citenamefont {Gharibyan}, \citenamefont {Gregorich}, \citenamefont {Kolomensky}, \citenamefont {Sammis}, \citenamefont {Scielzo}, \citenamefont {Shusterman},\ and\ \citenamefont {Thomas}}]{PhysRevC.109.055501}%
  \BibitemOpen
  \bibfield  {author} {\bibinfo {author} {\bibfnamefont {E.~B.}\ \bibnamefont {Norman}}, \bibinfo {author} {\bibfnamefont {A.}~\bibnamefont {Drobizhev}}, \bibinfo {author} {\bibfnamefont {N.}~\bibnamefont {Gharibyan}}, \bibinfo {author} {\bibfnamefont {K.~E.}\ \bibnamefont {Gregorich}}, \bibinfo {author} {\bibfnamefont {Y.~G.}\ \bibnamefont {Kolomensky}}, \bibinfo {author} {\bibfnamefont {B.~N.}\ \bibnamefont {Sammis}}, \bibinfo {author} {\bibfnamefont {N.~D.}\ \bibnamefont {Scielzo}}, \bibinfo {author} {\bibfnamefont {J.~A.}\ \bibnamefont {Shusterman}}, \ and\ \bibinfo {author} {\bibfnamefont {K.~J.}\ \bibnamefont {Thomas}},\ }\href {\doibase 10.1103/PhysRevC.109.055501} {\bibfield  {journal} {\bibinfo  {journal} {Phys. Rev. C}\ }\textbf {\bibinfo {volume} {109}},\ \bibinfo {pages} {055501} (\bibinfo {year} {2024})}\BibitemShut {NoStop}%
\bibitem [{\citenamefont {Collar}\ and\ \citenamefont {Yoon}(2023)}]{PhysRevC.108.L021602}%
  \BibitemOpen
  \bibfield  {author} {\bibinfo {author} {\bibfnamefont {J.~I.}\ \bibnamefont {Collar}}\ and\ \bibinfo {author} {\bibfnamefont {S.~G.}\ \bibnamefont {Yoon}},\ }\href {\doibase 10.1103/PhysRevC.108.L021602} {\bibfield  {journal} {\bibinfo  {journal} {Phys. Rev. C}\ }\textbf {\bibinfo {volume} {108}},\ \bibinfo {pages} {L021602} (\bibinfo {year} {2023})}\BibitemShut {NoStop}%
\bibitem [{\citenamefont {Hampel}\ and\ \citenamefont {Remsberg}(1985)}]{PhysRevC.31.666}%
  \BibitemOpen
  \bibfield  {author} {\bibinfo {author} {\bibfnamefont {W.}~\bibnamefont {Hampel}}\ and\ \bibinfo {author} {\bibfnamefont {L.~P.}\ \bibnamefont {Remsberg}},\ }\href {\doibase 10.1103/PhysRevC.31.666} {\bibfield  {journal} {\bibinfo  {journal} {Phys. Rev. C}\ }\textbf {\bibinfo {volume} {31}},\ \bibinfo {pages} {666} (\bibinfo {year} {1985})}\BibitemShut {NoStop}%
\bibitem [{\citenamefont {Alanssari}\ \emph {et~al.}(2016)\citenamefont {Alanssari}, \citenamefont {Frekers}, \citenamefont {Eronen}, \citenamefont {Canete}, \citenamefont {Hakala}, \citenamefont {Holl}, \citenamefont {Jokinen}, \citenamefont {Kankainen}, \citenamefont {Koponen}, \citenamefont {Moore}, \citenamefont {Nesterenko}, \citenamefont {Pohjalainen}, \citenamefont {Reinikainen}, \citenamefont {Rinta-Antila},\ and\ \citenamefont {Voss}}]{ALANSSARI20161}%
  \BibitemOpen
  \bibfield  {author} {\bibinfo {author} {\bibfnamefont {M.}~\bibnamefont {Alanssari}}, \bibinfo {author} {\bibfnamefont {D.}~\bibnamefont {Frekers}}, \bibinfo {author} {\bibfnamefont {T.}~\bibnamefont {Eronen}}, \bibinfo {author} {\bibfnamefont {L.}~\bibnamefont {Canete}}, \bibinfo {author} {\bibfnamefont {J.}~\bibnamefont {Hakala}}, \bibinfo {author} {\bibfnamefont {M.}~\bibnamefont {Holl}}, \bibinfo {author} {\bibfnamefont {A.}~\bibnamefont {Jokinen}}, \bibinfo {author} {\bibfnamefont {A.}~\bibnamefont {Kankainen}}, \bibinfo {author} {\bibfnamefont {J.}~\bibnamefont {Koponen}}, \bibinfo {author} {\bibfnamefont {I.}~\bibnamefont {Moore}}, \bibinfo {author} {\bibfnamefont {D.}~\bibnamefont {Nesterenko}}, \bibinfo {author} {\bibfnamefont {I.}~\bibnamefont {Pohjalainen}}, \bibinfo {author} {\bibfnamefont {J.}~\bibnamefont {Reinikainen}}, \bibinfo {author} {\bibfnamefont {S.}~\bibnamefont {Rinta-Antila}}, \ and\ \bibinfo {author} {\bibfnamefont {A.}~\bibnamefont {Voss}},\ }\href {\doibase
  https://doi.org/10.1016/j.ijms.2016.05.019} {\bibfield  {journal} {\bibinfo  {journal} {International Journal of Mass Spectrometry}\ }\textbf {\bibinfo {volume} {406}},\ \bibinfo {pages} {1} (\bibinfo {year} {2016})}\BibitemShut {NoStop}%
\bibitem [{\citenamefont {Cadeddu}\ \emph {et~al.}(2026)\citenamefont {Cadeddu}, \citenamefont {Cargioli}, \citenamefont {Dordei}, \citenamefont {Ferro}, \citenamefont {Giunti},\ and\ \citenamefont {Pitzalis}}]{Cadeddu2026}%
  \BibitemOpen
  \bibfield  {author} {\bibinfo {author} {\bibfnamefont {M.}~\bibnamefont {Cadeddu}}, \bibinfo {author} {\bibfnamefont {N.}~\bibnamefont {Cargioli}}, \bibinfo {author} {\bibfnamefont {F.}~\bibnamefont {Dordei}}, \bibinfo {author} {\bibfnamefont {L.}~\bibnamefont {Ferro}}, \bibinfo {author} {\bibfnamefont {C.}~\bibnamefont {Giunti}}, \ and\ \bibinfo {author} {\bibfnamefont {M.}~\bibnamefont {Pitzalis}},\ }\href {\doibase 10.1103/5pvf-mcr1} {\bibfield  {journal} {\bibinfo  {journal} {Phys. Rev. Lett.}\ }\textbf {\bibinfo {volume} {137}},\ \bibinfo {pages} {131805} (\bibinfo {year} {2026})}\BibitemShut {NoStop}%
\bibitem [{\citenamefont {{Mougeot}}(2018)}]{Mougeot_2018AppRI.134..225M}%
  \BibitemOpen
  \bibfield  {author} {\bibinfo {author} {\bibfnamefont {X.}~\bibnamefont {{Mougeot}}},\ }\href {\doibase 10.1016/j.apradiso.2017.07.027} {\bibfield  {journal} {\bibinfo  {journal} {Applied Radiation and Isotopes}\ }\textbf {\bibinfo {volume} {134}},\ \bibinfo {pages} {225} (\bibinfo {year} {2018})}\BibitemShut {NoStop}%
\bibitem [{\citenamefont {{Firestone}}\ \emph {et~al.}(1978)\citenamefont {{Firestone}}, \citenamefont {{McHarris}},\ and\ \citenamefont {{Holstein}}}]{Firestone_1978PhRvC..18.2719F}%
  \BibitemOpen
  \bibfield  {author} {\bibinfo {author} {\bibfnamefont {R.~B.}\ \bibnamefont {{Firestone}}}, \bibinfo {author} {\bibfnamefont {W.~C.}\ \bibnamefont {{McHarris}}}, \ and\ \bibinfo {author} {\bibfnamefont {B.~R.}\ \bibnamefont {{Holstein}}},\ }\href {\doibase 10.1103/PhysRevC.18.2719} {\bibfield  {journal} {\bibinfo  {journal} {\prc}\ }\textbf {\bibinfo {volume} {18}},\ \bibinfo {pages} {2719} (\bibinfo {year} {1978})}\BibitemShut {NoStop}%
\bibitem [{\citenamefont {Hardy}\ \emph {et~al.}(2018)\citenamefont {Hardy}, \citenamefont {Nica}, \citenamefont {Iacob},\ and\ \citenamefont {Trzhaskovskaya}}]{Hardy:2017exb}%
  \BibitemOpen
  \bibfield  {author} {\bibinfo {author} {\bibfnamefont {J.~C.}\ \bibnamefont {Hardy}}, \bibinfo {author} {\bibfnamefont {N.}~\bibnamefont {Nica}}, \bibinfo {author} {\bibfnamefont {V.~E.}\ \bibnamefont {Iacob}}, \ and\ \bibinfo {author} {\bibfnamefont {M.~B.}\ \bibnamefont {Trzhaskovskaya}},\ }\href {\doibase 10.1016/j.apradiso.2017.05.021} {\bibfield  {journal} {\bibinfo  {journal} {Appl. Radiat. Isot.}\ }\textbf {\bibinfo {volume} {134}},\ \bibinfo {pages} {406} (\bibinfo {year} {2018})},\ \Eprint {http://arxiv.org/abs/1705.11148} {arXiv:1705.11148 [nucl-ex]} \BibitemShut {NoStop}%
\bibitem [{\citenamefont {Basilico}\ \emph {et~al.}(2023)\citenamefont {Basilico} \emph {et~al.}}]{Borexino_PhysRevD.108.102005}%
  \BibitemOpen
  \bibfield  {author} {\bibinfo {author} {\bibfnamefont {D.}~\bibnamefont {Basilico}} \emph {et~al.} (\bibinfo {collaboration} {Borexino Collaboration}),\ }\href {\doibase 10.1103/PhysRevD.108.102005} {\bibfield  {journal} {\bibinfo  {journal} {Phys. Rev. D}\ }\textbf {\bibinfo {volume} {108}},\ \bibinfo {pages} {102005} (\bibinfo {year} {2023})}\BibitemShut {NoStop}%
\bibitem [{\citenamefont {Agostini}\ \emph {et~al.}(2018)\citenamefont {Agostini} \emph {et~al.}}]{Borexino2018}%
  \BibitemOpen
  \bibfield  {author} {\bibinfo {author} {\bibfnamefont {M.}~\bibnamefont {Agostini}} \emph {et~al.} (\bibinfo {collaboration} {Borexino Collaboration}),\ }\href {\doibase https://doi.org/10.1038/s41586-018-0624-y} {\bibfield  {journal} {\bibinfo  {journal} {Nature}\ }\textbf {\bibinfo {volume} {562}},\ \bibinfo {pages} {505} (\bibinfo {year} {2018})}\BibitemShut {NoStop}%
\bibitem [{\citenamefont {Agostini}\ \emph {et~al.}(2020)\citenamefont {Agostini} \emph {et~al.}}]{BOREXINO:2020aww}%
  \BibitemOpen
  \bibfield  {author} {\bibinfo {author} {\bibfnamefont {M.}~\bibnamefont {Agostini}} \emph {et~al.} (\bibinfo {collaboration} {BOREXINO}),\ }\href {\doibase 10.1038/s41586-020-2934-0} {\bibfield  {journal} {\bibinfo  {journal} {Nature}\ }\textbf {\bibinfo {volume} {587}},\ \bibinfo {pages} {577} (\bibinfo {year} {2020})},\ \Eprint {http://arxiv.org/abs/2006.15115} {arXiv:2006.15115 [hep-ex]} \BibitemShut {NoStop}%
\bibitem [{\citenamefont {Aharmim}\ \emph {et~al.}(2013)\citenamefont {Aharmim} \emph {et~al.}}]{SNO:2011hxd}%
  \BibitemOpen
  \bibfield  {author} {\bibinfo {author} {\bibfnamefont {B.}~\bibnamefont {Aharmim}} \emph {et~al.} (\bibinfo {collaboration} {SNO}),\ }\href {\doibase 10.1103/PhysRevC.88.025501} {\bibfield  {journal} {\bibinfo  {journal} {Phys. Rev. C}\ }\textbf {\bibinfo {volume} {88}},\ \bibinfo {pages} {025501} (\bibinfo {year} {2013})},\ \Eprint {http://arxiv.org/abs/1109.0763} {arXiv:1109.0763 [nucl-ex]} \BibitemShut {NoStop}%
\bibitem [{\citenamefont {Yankelevich}\ \emph {et~al.}(2026)\citenamefont {Yankelevich} \emph {et~al.}}]{Super-Kamiokande:2025hvx}%
  \BibitemOpen
  \bibfield  {author} {\bibinfo {author} {\bibfnamefont {A.}~\bibnamefont {Yankelevich}} \emph {et~al.} (\bibinfo {collaboration} {Super-Kamiokande}),\ }\href {\doibase 10.1103/yz79-h758} {\bibfield  {journal} {\bibinfo  {journal} {Phys. Rev. D}\ }\textbf {\bibinfo {volume} {113}},\ \bibinfo {pages} {112001} (\bibinfo {year} {2026})},\ \Eprint {http://arxiv.org/abs/2512.19887} {arXiv:2512.19887 [hep-ex]} \BibitemShut {NoStop}%
\bibitem [{\citenamefont {Abe}\ \emph {et~al.}(2024)\citenamefont {Abe} \emph {et~al.}}]{Super-Kamiokande:2023jbt}%
  \BibitemOpen
  \bibfield  {author} {\bibinfo {author} {\bibfnamefont {K.}~\bibnamefont {Abe}} \emph {et~al.} (\bibinfo {collaboration} {Super-Kamiokande}),\ }\href {\doibase 10.1103/PhysRevD.109.092001} {\bibfield  {journal} {\bibinfo  {journal} {Phys. Rev. D}\ }\textbf {\bibinfo {volume} {109}},\ \bibinfo {pages} {092001} (\bibinfo {year} {2024})},\ \Eprint {http://arxiv.org/abs/2312.12907} {arXiv:2312.12907 [hep-ex]} \BibitemShut {NoStop}%
\bibitem [{\citenamefont {'t~Hooft}(1971)}]{tHooft:1971ucy}%
  \BibitemOpen
  \bibfield  {author} {\bibinfo {author} {\bibfnamefont {G.}~\bibnamefont {'t~Hooft}},\ }\href {\doibase 10.1016/0370-2693(71)90050-5} {\bibfield  {journal} {\bibinfo  {journal} {Phys. Lett. B}\ }\textbf {\bibinfo {volume} {37}},\ \bibinfo {pages} {195} (\bibinfo {year} {1971})}\BibitemShut {NoStop}%
\bibitem [{\citenamefont {Butler}\ \emph {et~al.}(2001)\citenamefont {Butler}, \citenamefont {Chen},\ and\ \citenamefont {Kong}}]{Butler:2000zp}%
  \BibitemOpen
  \bibfield  {author} {\bibinfo {author} {\bibfnamefont {M.}~\bibnamefont {Butler}}, \bibinfo {author} {\bibfnamefont {J.-W.}\ \bibnamefont {Chen}}, \ and\ \bibinfo {author} {\bibfnamefont {X.}~\bibnamefont {Kong}},\ }\href {\doibase 10.1103/PhysRevC.63.035501} {\bibfield  {journal} {\bibinfo  {journal} {Phys. Rev. C}\ }\textbf {\bibinfo {volume} {63}},\ \bibinfo {pages} {035501} (\bibinfo {year} {2001})},\ \Eprint {http://arxiv.org/abs/nucl-th/0008032} {arXiv:nucl-th/0008032} \BibitemShut {NoStop}%
\bibitem [{\citenamefont {Ahmad}\ \emph {et~al.}(2002)\citenamefont {Ahmad} \emph {et~al.}}]{SNO:2002tuh}%
  \BibitemOpen
  \bibfield  {author} {\bibinfo {author} {\bibfnamefont {Q.~R.}\ \bibnamefont {Ahmad}} \emph {et~al.} (\bibinfo {collaboration} {SNO}),\ }\href {\doibase 10.1103/PhysRevLett.89.011301} {\bibfield  {journal} {\bibinfo  {journal} {Phys. Rev. Lett.}\ }\textbf {\bibinfo {volume} {89}},\ \bibinfo {pages} {011301} (\bibinfo {year} {2002})},\ \Eprint {http://arxiv.org/abs/nucl-ex/0204008} {arXiv:nucl-ex/0204008} \BibitemShut {NoStop}%
\bibitem [{\citenamefont {Abdurashitov}\ \emph {et~al.}(2009)\citenamefont {Abdurashitov} \emph {et~al.}}]{SAGE:2009eeu}%
  \BibitemOpen
  \bibfield  {author} {\bibinfo {author} {\bibfnamefont {J.~N.}\ \bibnamefont {Abdurashitov}} \emph {et~al.} (\bibinfo {collaboration} {SAGE}),\ }\href {\doibase 10.1103/PhysRevC.80.015807} {\bibfield  {journal} {\bibinfo  {journal} {Phys. Rev. C}\ }\textbf {\bibinfo {volume} {80}},\ \bibinfo {pages} {015807} (\bibinfo {year} {2009})},\ \Eprint {http://arxiv.org/abs/0901.2200} {arXiv:0901.2200 [nucl-ex]} \BibitemShut {NoStop}%
\bibitem [{\citenamefont {Cleveland}\ \emph {et~al.}(1995)\citenamefont {Cleveland}, \citenamefont {Daily}, \citenamefont {Davis}, \citenamefont {Distel}, \citenamefont {Lande}, \citenamefont {Lee}, \citenamefont {Wildenhain},\ and\ \citenamefont {Ullman}}]{Cleveland:1994er}%
  \BibitemOpen
  \bibfield  {author} {\bibinfo {author} {\bibfnamefont {B.~T.}\ \bibnamefont {Cleveland}}, \bibinfo {author} {\bibfnamefont {T.}~\bibnamefont {Daily}}, \bibinfo {author} {\bibfnamefont {R.}~\bibnamefont {Davis}, \bibfnamefont {Jr.}}, \bibinfo {author} {\bibfnamefont {J.}~\bibnamefont {Distel}}, \bibinfo {author} {\bibfnamefont {K.}~\bibnamefont {Lande}}, \bibinfo {author} {\bibfnamefont {C.~K.}\ \bibnamefont {Lee}}, \bibinfo {author} {\bibfnamefont {P.}~\bibnamefont {Wildenhain}}, \ and\ \bibinfo {author} {\bibfnamefont {J.}~\bibnamefont {Ullman}},\ }\href {\doibase 10.1016/0920-5632(94)00732-B} {\bibfield  {journal} {\bibinfo  {journal} {Nucl. Phys. B Proc. Suppl.}\ }\textbf {\bibinfo {volume} {38}},\ \bibinfo {pages} {47} (\bibinfo {year} {1995})}\BibitemShut {NoStop}%
\bibitem [{\citenamefont {Haxton}\ and\ \citenamefont {Rule}(2025)}]{Haxton:2025hye}%
  \BibitemOpen
  \bibfield  {author} {\bibinfo {author} {\bibfnamefont {W.~C.}\ \bibnamefont {Haxton}}\ and\ \bibinfo {author} {\bibfnamefont {E.}~\bibnamefont {Rule}},\ }\href {\doibase 10.1016/j.physletb.2025.139259} {\bibfield  {journal} {\bibinfo  {journal} {Phys. Lett. B}\ }\textbf {\bibinfo {volume} {861}},\ \bibinfo {pages} {139259} (\bibinfo {year} {2025})},\ \Eprint {http://arxiv.org/abs/2501.03528} {arXiv:2501.03528 [nucl-ex]} \BibitemShut {NoStop}%
\bibitem [{\citenamefont {Esteban}\ \emph {et~al.}(2017)\citenamefont {Esteban}, \citenamefont {Gonzalez-Garcia}, \citenamefont {Maltoni}, \citenamefont {Martinez-Soler},\ and\ \citenamefont {Schwetz}}]{Esteban:2016qun}%
  \BibitemOpen
  \bibfield  {author} {\bibinfo {author} {\bibfnamefont {I.}~\bibnamefont {Esteban}}, \bibinfo {author} {\bibfnamefont {M.~C.}\ \bibnamefont {Gonzalez-Garcia}}, \bibinfo {author} {\bibfnamefont {M.}~\bibnamefont {Maltoni}}, \bibinfo {author} {\bibfnamefont {I.}~\bibnamefont {Martinez-Soler}}, \ and\ \bibinfo {author} {\bibfnamefont {T.}~\bibnamefont {Schwetz}},\ }\href {\doibase 10.1007/JHEP01(2017)087} {\bibfield  {journal} {\bibinfo  {journal} {JHEP}\ }\textbf {\bibinfo {volume} {01}},\ \bibinfo {pages} {087} (\bibinfo {year} {2017})},\ \Eprint {http://arxiv.org/abs/1611.01514} {arXiv:1611.01514 [hep-ph]} \BibitemShut {NoStop}%
\bibitem [{\citenamefont {Adam}\ \emph {et~al.}(2026)\citenamefont {Adam} \emph {et~al.}}]{thejunocollaboration2026measurementneutrinooscillationparameters}%
  \BibitemOpen
  \bibfield  {author} {\bibinfo {author} {\bibfnamefont {T.}~\bibnamefont {Adam}} \emph {et~al.} (\bibinfo {collaboration} {The JUNO Collaboration}),\ }\href {https://arxiv.org/abs/2609.37895} {\enquote {\bibinfo {title} {{Measurement of Neutrino Oscillation Parameters at JUNO and Indication of the Neutrino Mass Ordering}},}\ } (\bibinfo {year} {2026}),\ \Eprint {http://arxiv.org/abs/2609.37895} {arXiv:2609.37895 [hep-ex]} \BibitemShut {NoStop}%
\bibitem [{\citenamefont {Herrera}\ and\ \citenamefont {Serenelli}(2023)}]{Herrera2023}%
  \BibitemOpen
  \bibfield  {author} {\bibinfo {author} {\bibfnamefont {Y.}~\bibnamefont {Herrera}}\ and\ \bibinfo {author} {\bibfnamefont {A.}~\bibnamefont {Serenelli}},\ }\href@noop {} {\enquote {\bibinfo {title} {{Standard Solar Models B23 / SF-III}},}\ } (\bibinfo {year} {2023}),\ \bibinfo {note} {{https://zenodo.org/records/10822316}}\BibitemShut {NoStop}%
\bibitem [{\citenamefont {Bahcall}(1989)}]{Bahcall:1989ks}%
  \BibitemOpen
  \bibfield  {author} {\bibinfo {author} {\bibfnamefont {J.~N.}\ \bibnamefont {Bahcall}},\ }\href@noop {} {\emph {\bibinfo {title} {{NEUTRINO ASTROPHYSICS}}}}\ (\bibinfo  {publisher} {Cambridge University Press},\ \bibinfo {year} {1989})\BibitemShut {NoStop}%
\bibitem [{\citenamefont {Winter}\ \emph {et~al.}(2006)\citenamefont {Winter}, \citenamefont {Freedman}, \citenamefont {Rehm},\ and\ \citenamefont {Schiffer}}]{Winter:2004kf}%
  \BibitemOpen
  \bibfield  {author} {\bibinfo {author} {\bibfnamefont {W.~T.}\ \bibnamefont {Winter}}, \bibinfo {author} {\bibfnamefont {S.~J.}\ \bibnamefont {Freedman}}, \bibinfo {author} {\bibfnamefont {K.~E.}\ \bibnamefont {Rehm}}, \ and\ \bibinfo {author} {\bibfnamefont {J.~P.}\ \bibnamefont {Schiffer}},\ }\href {\doibase 10.1103/PhysRevC.73.025503} {\bibfield  {journal} {\bibinfo  {journal} {Phys. Rev. C}\ }\textbf {\bibinfo {volume} {73}},\ \bibinfo {pages} {025503} (\bibinfo {year} {2006})},\ \Eprint {http://arxiv.org/abs/nucl-ex/0406019} {arXiv:nucl-ex/0406019} \BibitemShut {NoStop}%
\bibitem [{\citenamefont {Acharya}\ \emph {et~al.}(2025)\citenamefont {Acharya} \emph {et~al.}}]{Acharya:2024lke}%
  \BibitemOpen
  \bibfield  {author} {\bibinfo {author} {\bibfnamefont {B.}~\bibnamefont {Acharya}} \emph {et~al.},\ }\href {\doibase 10.1103/8lm7-gs18} {\bibfield  {journal} {\bibinfo  {journal} {Rev. Mod. Phys.}\ }\textbf {\bibinfo {volume} {97}},\ \bibinfo {pages} {035002} (\bibinfo {year} {2025})},\ \Eprint {http://arxiv.org/abs/2405.06470} {arXiv:2405.06470 [astro-ph.SR]} \BibitemShut {NoStop}%
\bibitem [{\citenamefont {Longfellow}\ \emph {et~al.}(2023)\citenamefont {Longfellow} \emph {et~al.}}]{Longfellow:2023hoj}%
  \BibitemOpen
  \bibfield  {author} {\bibinfo {author} {\bibfnamefont {B.}~\bibnamefont {Longfellow}} \emph {et~al.},\ }\href {\doibase 10.1103/PhysRevC.107.L032801} {\bibfield  {journal} {\bibinfo  {journal} {Phys. Rev. C}\ }\textbf {\bibinfo {volume} {107}},\ \bibinfo {pages} {L032801} (\bibinfo {year} {2023})}\BibitemShut {NoStop}%
\bibitem [{\citenamefont {Bahcall}\ and\ \citenamefont {Krastev}(1996)}]{Bahcall:1995rs}%
  \BibitemOpen
  \bibfield  {author} {\bibinfo {author} {\bibfnamefont {J.~N.}\ \bibnamefont {Bahcall}}\ and\ \bibinfo {author} {\bibfnamefont {P.~I.}\ \bibnamefont {Krastev}},\ }\href {\doibase 10.1103/PhysRevD.53.4211} {\bibfield  {journal} {\bibinfo  {journal} {Phys. Rev. D}\ }\textbf {\bibinfo {volume} {53}},\ \bibinfo {pages} {4211} (\bibinfo {year} {1996})},\ \Eprint {http://arxiv.org/abs/hep-ph/9512378} {arXiv:hep-ph/9512378} \BibitemShut {NoStop}%
\bibitem [{\citenamefont {Heeger}\ and\ \citenamefont {Robertson}(1996)}]{Heeger:1996se}%
  \BibitemOpen
  \bibfield  {author} {\bibinfo {author} {\bibfnamefont {K.~M.}\ \bibnamefont {Heeger}}\ and\ \bibinfo {author} {\bibfnamefont {R.~G.~H.}\ \bibnamefont {Robertson}},\ }\href {\doibase 10.1103/PhysRevLett.77.3720} {\bibfield  {journal} {\bibinfo  {journal} {Phys. Rev. Lett.}\ }\textbf {\bibinfo {volume} {77}},\ \bibinfo {pages} {3720} (\bibinfo {year} {1996})},\ \Eprint {http://arxiv.org/abs/nucl-th/9610030} {arXiv:nucl-th/9610030} \BibitemShut {NoStop}%
\bibitem [{\citenamefont {Gonzalez-Garcia}\ \emph {et~al.}(2024)\citenamefont {Gonzalez-Garcia}, \citenamefont {Maltoni}, \citenamefont {Pinheiro},\ and\ \citenamefont {Serenelli}}]{Gonzalez-Garcia:2023kva}%
  \BibitemOpen
  \bibfield  {author} {\bibinfo {author} {\bibfnamefont {M.~C.}\ \bibnamefont {Gonzalez-Garcia}}, \bibinfo {author} {\bibfnamefont {M.}~\bibnamefont {Maltoni}}, \bibinfo {author} {\bibfnamefont {J.~P.}\ \bibnamefont {Pinheiro}}, \ and\ \bibinfo {author} {\bibfnamefont {A.~M.}\ \bibnamefont {Serenelli}},\ }\href {\doibase 10.1007/JHEP02(2024)064} {\bibfield  {journal} {\bibinfo  {journal} {JHEP}\ }\textbf {\bibinfo {volume} {02}},\ \bibinfo {pages} {064} (\bibinfo {year} {2024})},\ \Eprint {http://arxiv.org/abs/2311.16226} {arXiv:2311.16226 [hep-ph]} \BibitemShut {NoStop}%
\bibitem [{\citenamefont {Bahcall}(1997)}]{Bahcall:1997eg}%
  \BibitemOpen
  \bibfield  {author} {\bibinfo {author} {\bibfnamefont {J.~N.}\ \bibnamefont {Bahcall}},\ }\href {\doibase 10.1103/PhysRevC.56.3391} {\bibfield  {journal} {\bibinfo  {journal} {Phys. Rev. C}\ }\textbf {\bibinfo {volume} {56}},\ \bibinfo {pages} {3391} (\bibinfo {year} {1997})},\ \Eprint {http://arxiv.org/abs/hep-ph/9710491} {arXiv:hep-ph/9710491} \BibitemShut {NoStop}%
\bibitem [{\citenamefont {Bahcall}\ and\ \citenamefont {Pe\~{n}a Garay}(2004)}]{BahcallNJP2004}%
  \BibitemOpen
  \bibfield  {author} {\bibinfo {author} {\bibfnamefont {J.~N.}\ \bibnamefont {Bahcall}}\ and\ \bibinfo {author} {\bibfnamefont {C.}~\bibnamefont {Pe\~{n}a Garay}},\ }\href {\doibase 10.1088/1367-2630/6/1/063} {\bibfield  {journal} {\bibinfo  {journal} {New Journal of Physics}\ }\textbf {\bibinfo {volume} {6}},\ \bibinfo {pages} {63} (\bibinfo {year} {2004})}\BibitemShut {NoStop}%
\bibitem [{\citenamefont {An}\ \emph {et~al.}(2023)\citenamefont {An} \emph {et~al.}}]{PhysRevLett.130.161802}%
  \BibitemOpen
  \bibfield  {author} {\bibinfo {author} {\bibfnamefont {F.~P.}\ \bibnamefont {An}} \emph {et~al.} (\bibinfo {collaboration} {Daya Bay Collaboration}),\ }\href {\doibase 10.1103/PhysRevLett.130.161802} {\bibfield  {journal} {\bibinfo  {journal} {Phys. Rev. Lett.}\ }\textbf {\bibinfo {volume} {130}},\ \bibinfo {pages} {161802} (\bibinfo {year} {2023})}\BibitemShut {NoStop}%
\bibitem [{\citenamefont {{Magg}}\ \emph {et~al.}(2022)\citenamefont {{Magg}}, \citenamefont {{Bergemann}}, \citenamefont {{Serenelli}}, \citenamefont {{Bautista}}, \citenamefont {{Plez}}, \citenamefont {{Heiter}}, \citenamefont {{Gerber}}, \citenamefont {{Ludwig}}, \citenamefont {{Basu}}, \citenamefont {{Ferguson}}, \citenamefont {{Gallego}}, \citenamefont {{Gamrath}}, \citenamefont {{Palmeri}},\ and\ \citenamefont {{Quinet}}}]{Magg2022}%
  \BibitemOpen
  \bibfield  {author} {\bibinfo {author} {\bibfnamefont {E.}~\bibnamefont {{Magg}}}, \bibinfo {author} {\bibfnamefont {M.}~\bibnamefont {{Bergemann}}}, \bibinfo {author} {\bibfnamefont {A.}~\bibnamefont {{Serenelli}}}, \bibinfo {author} {\bibfnamefont {M.}~\bibnamefont {{Bautista}}}, \bibinfo {author} {\bibfnamefont {B.}~\bibnamefont {{Plez}}}, \bibinfo {author} {\bibfnamefont {U.}~\bibnamefont {{Heiter}}}, \bibinfo {author} {\bibfnamefont {J.~M.}\ \bibnamefont {{Gerber}}}, \bibinfo {author} {\bibfnamefont {H.-G.}\ \bibnamefont {{Ludwig}}}, \bibinfo {author} {\bibfnamefont {S.}~\bibnamefont {{Basu}}}, \bibinfo {author} {\bibfnamefont {J.~W.}\ \bibnamefont {{Ferguson}}}, \bibinfo {author} {\bibfnamefont {H.~C.}\ \bibnamefont {{Gallego}}}, \bibinfo {author} {\bibfnamefont {S.}~\bibnamefont {{Gamrath}}}, \bibinfo {author} {\bibfnamefont {P.}~\bibnamefont {{Palmeri}}}, \ and\ \bibinfo {author} {\bibfnamefont {P.}~\bibnamefont {{Quinet}}},\ }\href {\doibase 10.1051/0004-6361/202142971} {\bibfield  {journal}
  {\bibinfo  {journal} {Astron. Astrophys.}\ }\textbf {\bibinfo {volume} {661}},\ \bibinfo {pages} {A140} (\bibinfo {year} {2022})},\ \Eprint {http://arxiv.org/abs/2203.02255} {arXiv:2203.02255 [astro-ph.SR]} \BibitemShut {NoStop}%
\end{thebibliography}%

\end{document}